\documentstyle[aps,preprint,prb]{revtex}
\def\({\begin{equation}}
\def\){\end{equation}}
\begin{document}                
\title{Finite-Temperature Transition into a Power-Law Spin Phase with an
Extensive Zero-Point Entropy}
\author{P. Chandra}
\address{NEC Research Institute, 4 Independence Way, Princeton, NJ 08540.}
\author{P. Coleman and L.B. Ioffe}
\address{Serin Physics Laboratory, Rutgers University, P.O. Box 849,
Piscataway, NJ 08854.}
\maketitle
\begin{abstract}
We introduce an $xy$ generalization of the frustrated
Ising model on a triangular lattice.  The presence of continuous
degrees of freedom stabilizes a {\em finite-temperature} spin
state
with {\em  power-law} discrete spin correlations and
an extensive zero-point entropy.  In this phase,
the unquenched degrees of freedom can be described by a fluctuating
surface with logarithmic height correlations.  Finite-size
Monte Carlo simulations have been used to characterize
the exponents of the transition and the dynamics of the low-temperature
phase.
\end{abstract}

\narrowtext

\section{Introduction}

The possibility of spin systems with a residual zero-point entropy has existed
since Wannier's study of  the two-dimensional Ising triangular
antiferromagnet; \cite{wannier}   here the frustration leads to a
zero-temperature ($T=0$) critical point
characterized by an extensive entropy $S_0 = 0.338$.
The associated unquenched degrees of freedom can be described by
a fluctuating surface with logarithmic height correlations;\cite{steph,bh}
in discrete spin models the weak nature of the long-range entropic
interactions between height defects
forbids the existence of this critical phase at $T \neq 0$.
Can a phase transition
into a state with a zero-point entropy exist at finite temperatures?
This issue was first raised by Anderson\cite{pwa0} in his study
of spinels, where he noted that a [111] projection
of this structure corresponds to a planar kagom\'e net.
Anderson suggested that the ground-state manifold of Heisenberg spins
on such frustrated lattices would be highly degenerate,
and thus that long-range magnetic order would be suppressed
due to the residual entropy at finite temperatures.\cite{pwa0}
The resulting phase would have {\em  power-law} discrete spin correlations
at finite temperatures in contrast to the exponential and constant Ising
correlations associated with a paramagnet and an ordered phase;
it is thus natural to call this state a ``classical spin liquid''.

There now exists an experimental realization of Anderson's model:
the magnetoplumbite $SrCr_{8-x}Ga_{4+x}O_{19}$ ($SCGO(x)$), whose magnetic
properties are attributed to planes of antiferromagnetically-coupled
chromium atoms on a kagom\'e lattice.\cite{ob}
Though inelastic neutron scattering\cite{neutrons} and specific
heat\cite{art} measurements on $SCGO(x)$ are consistent
with antiferromagnetic order, the observed spin correlation
length is only a few lattice spacings.\cite{neutrons} There
exist many other anomalous magnetic
materials that, like $SCGO(x)$, do not
display conventional long-range spin order.\cite{houches,art2}
Most of these experimental systems
have physical structures that impose strong constraints on the
low-temperature configurations of their continuous spins, allowing them to be
identified with the ground-states of an associated discrete model on the same
lattice.  Following the spirit of Anderson's original suggestion,\cite{pwa0}
we show that an interplay between continuous and
discrete degrees of freedom can enhance the
defect-defect interaction in the associated fluctuating surface, thus
stabilizing a finite-temperature critical point associated with a zero-point
entropy.

\section{The Discrete Model}

The two-dimensional Ising antiferromagnet with nearest-neighbor interactions on
the triangular lattice is the simplest frustrated spin system;\cite{wannier}
unlike
its ferromagnetic counterpart it does not order even at zero
temperature but has {\em  power-law} spin correlations in
contrast to a conventional paramagnet.\cite{steph}  The geometry of this
lattice
does not permit the minimization of each bond energetically and thus
even at $T=0$ there exist parallel {\em  and} antiparallel nearest-neighbor
spin pairs which we denote as antiferromagnetic (AFM) and ferromagnetic
(FM) bonds for convenience.  The energy of each triangle is minimized
if it contains one and only one ferromagnetic bond.  The number
of spin configurations that satisfy this plaquette constraint scales
exponentially with the site number; the resulting $T=0$ manifold
has an {\em  extensive} entropy and can be fully explored from a given
starting state by flipping spins with equal numbers of AFM and FM
bonds (Figure 1).  If a spin with {\em  four} AFM bonds is reversed there
result
{\em  two} ``bad'' triangles that violate the constraint (Figure 2).
These two ``defects''  can be separated to arbitrary distance at
no additional cost, despite the fact that in doing so all the spins
along the line linking them are reversed (Figure 3).
The situation is similar to that in the $1d$
Ising ferromagnet where a flip of a single spin creates two ``bad'' bonds.
These defects can move freely along the chain by reversing all the spins
between them; since no more ``bad'' bonds appear in this process
their energy cost is independent of their relative positions.
In the triangular case the analogous separation of
two ``bad'' triangles does not lead to the formation of
more ``bad'' plaquettes, so that their energy cost is also
finite in the thermodynamic limit.
At $T=0$ the creation of these
defects is forbidden, there is no length-scale in the problem and the
spin correlations are power-law.  However at nonzero temperatures
defect formation is favored entropically, and the spin correlations
become exponential with a correlation length determined by the
defect density. We contrast this situation
with that of the  $2d$ Ising {\sl ferromagnet} where the minimum energy
cost associated with two defects at a separation $L$ scales like $E \sim
L^{1\over 2}$; here defect formation is suppressed until a finite temperature.

We can characterize each spin state by its spatial configuration of
antiferromagnetic bonds.  At $T=0$ the plaquette constraint ensures that
each triangle has exactly two AFM bonds; each spin state can thus be mapped
onto a unique tiling of the plane by rhombi which can also be
viewed as a planar projection of a surface in three-dimensions (Figure 4).
Quantitatively
this mapping\cite{bh} between the spins $\{s_i\}$ and the heights
$\{h_i\}$ associated with the surface is defined by the expression
\(
(h_i - h_j) = {1\over2} (3\vec{s}_i \cdot \vec {s}_j + 1)
\label{height}
\)
for sites $i$ and $j$ as shown in Figure 4.If the underlying spin configuration
satisfies the plaquette constraint, the sum of the height differences
as defined in (\ref{height}) around any closed path will be exactly
zero (Figure 5a) and the height is a uniquely defined function.
By contrast, a similar sum determined along a loop enclosing a
``bad'' triangle (see Figure 5b ) results in a {\em  finite}
height difference which, in analogy with conventional
elasticity theory,
we denote as the Burger's vector; the underlying spin configuration determines
its numerical value to be a multiple of six.  Defects in the
surface appear because the height is now a multi-valued function whose
branching points correspond to triangles that violate the constraint.

The surface representation provides a very convenient description of
the full ground-state manifold; in this language the spin flips associated with
a zero energy cost correspond to the addition/subtraction of elementary
cubes.  Thus the allowed $T=0$ spin configurations
are mapped onto a free fluctuating surface whose behavior at finite
temperatures and long length-scales is characterized by the free energy
density
\(
F = {1\over 2} \eta T (\nabla h)^2
\label{fenergy}
\)
where $\eta T$ is the stiffness associated with the surface
tension of the membrane.  Equation (\ref{fenergy}) leads to
logarithmic height correlations of the form
\(
\langle (h_i - h_j)^2 \rangle = {1\over \eta\pi} \ln {r_{ij}\over a}
\label{hlog}
\)
where $a$ is the cutoff imposed by the lattice.  Since the spins and the
heights on {\em  any} sites $i$ and $j$ are related by the expression
\(
s_i \cdot s_j = (-1)^{\{h_i - h_j\}}
\label{spinheight}
\)
the spin-spin correlation function at large distances
is power-law
\(
\langle s_i \cdot s_j \rangle = \exp - {\pi^2\over 2} \langle (h_i - h_j)^2
\rangle = \left( a \over r_{ij} \right)^{\pi\over 2 \eta}
\label{scfunction}
\)
which follows from the free energy density (\ref{fenergy}).  For the $T
\rightarrow 0$ Ising model
with nearest-neighbor interactions that we have considered above,
the correlation function (\ref{scfunction}) is known from the exact
solution\cite{steph,bh}
and results in a stiffness coefficient
\(
\eta_{Ising} = {\pi\over 9}
\label{etaising}
\)
for the associated free surface.
In principle, a generalized spin model on the triangular lattice could lead to
a different numerical value of $\eta$, which is determined by the
specific form of the spin-spin interaction.

The free energy density (\ref{fenergy}) leads to a defect energy
that scales logarithmically with system size at any finite temperature.
The resulting free energy associated with an isolated defect is
\(
F_{defect} = E_0 - 2T\left\{ \left({b^2\eta}\over 8\pi\right) - 1 \right\} \ln
L
\label{fdefect}
\)
where $E_0$ is the core energy and the last term in (\ref{fdefect})
is the entropy associated with the random location
of the defect on the surface. This defect free energy
is logarithmically
large if the coefficient of the temperature in (\ref{fdefect})
is positive; for the case here where the minimum Burger's vector
$b_{min} =6$, the critical value for $\eta$ is
\(
\eta_{c_1} = {8\pi\over b^2} = {2\pi\over 9}
\label{eta1}
\)
Binding of the surface defects\cite{kt} occurs for $\eta \ge \eta_{c_1}$,
and we note that $\eta_{Ising} < \eta_{c_1}$
consistent with the known presence of exponential spin correlations at finite
temperatures.

In the previous discussion we have implicitly assumed the irrelevance
of terms in the free energy density that favor a particular surface
orientation.  These contributions should be periodic in the height
and thus have a general form
\(
F_r = \sum_n V_n \cos 2\pi n h
\label{cos}
\)
where the most ``dangerous'' term is the case $n=1$ with
dimension $[L]^{2 - {\eta\over \pi}}$.  Thus for a stiffness
constant
larger than
\(
\eta_{c_2} = {\pi\over 2}
\label{eta2}
\)
the surface no longer fluctuates on long length-scales but
remains essentally flat.\cite{noz}

The possibility of a fluctuating defect-free surface at finite
temperatures then exists for values of the stiffness coefficient in the region
\(
\eta_{c_1} > \eta > \eta_{c_2}
\label{window}
\)
where defects are bound but the surface is not spontaneously oriented.
In this window (\ref{window}) the correlations of the discrete spins
are power-law in contrast to their exponential and constant behavior
for $\eta < \eta_{c_1}$ and $\eta > \eta_{c_2}$ respectively (see Figure 6).
We note that this possibility for spin liquid behavior exists because
the minimum Burger's vector associated with the underlying spin
model is $b_{min} = 6$, which is twice
that allowed for the pure surface problem.
The value of the stiffness constant associated with the nearest-neighbor Ising
model
is {\em  less} than the critical value required for defect-binding,
consistent with the {\em  absence} of power-law Ising spin correlations
at any finite temperature.  We would like to construct a generalization
of this model with a spin-spin interaction that leads to an
enhanced stiffness coefficient.  In this paper, we have introduced additional
continuous degrees of freedom and show that their interaction
with the Ising variables leads to the stability of a classical
spin liquid phase at finite temperatures.

\section{The Generalized Spin Model}

The simplest frustrated model that has {\em  both} Ising and continuous
degrees of freedom is described by
the generalized $xy$ Hamiltonian
\(
H = {1\over 2}\sum_{ij} f(\Theta_i - \Theta_j)
\label{ham}
\)
on a triangular lattice.  We specify the periodic function
$f(\Theta_i - \Theta_j)$ so that in the ground-state configurations
the spins are nearly collinear with angular orientations described
by
\(
\Theta_i = \pi \sigma_i +
\theta_i
\label{angles}
\)
where $\sigma_i$ and $\Theta_i$
are the Ising and the continuous variables respectively.
The angular deviation between spin orientations is small
on short length-scales
(i.e. $|{\theta}_i - {\theta}_j| << \pi$
where $i$ and $j$ are nearest-neighbor sites)
but can be substantial at large distances.
We tune the
spectrum of the continuous degrees of
freedom to
be very soft so as to maximize their interaction with the Ising variables.
Specifically we choose the spin-spin interaction
\(
f(\Theta) = z \left\{ \cos \Theta + {1\over 12} \cos^4 \Theta \right\}
            - (1-z) \left\{ \cos^2 \Theta - {1\over 2} \cos^4 \Theta\right\}
\label{int}
\)
which admits {\em  collinear} ground-state configurations for
$z < z_c$;
the coefficients have been selected
so that the first
non-vanishing
term in a small-angle expansion
\(
f(\Theta) \sim \beta \Theta^2 + \gamma \Theta^4
\label{int2}
\)
of (\ref{int}) is quartic (i.e. $\beta = 0$)
while simultaneously maximizing the value of $z_c$.
This specific form of the spin-spin interaction
(\ref{int}) leads to very flat free-energy wells with small
angular fluctuations described by $<\theta^2> \sim T^{1\over 2}$.
We thus expect the presence of these soft continuous
degrees of freedom to modify the behavior of the discrete variables
of the model.  More specifically we hope that they will lead
to a significant enhancement of the height stiffness and thus
to the possibility of a spin liquid phase (Figure 6).
We note that a simpler form of $f(\Theta)$ with
$\beta \neq 0$ in (\ref{int2}) results in $<\theta^2> \sim T$;
this leads to a numerically small increase\cite{mc} of the surface stiffness
that does not produce any interesting physical effects.\cite{lee}

At $z=0$ the ground-state of the Hamiltonian
specified by (\ref{ham}) and (\ref{int}) is a disordered nematic where the
$xy$ variables
are ordered but the Ising are not.
The finite $z$ term provides an interaction between
the discrete variables, and thus their behavior can be described
by a fluctuating surface as $T \rightarrow 0$.
For $z \neq 0$ there is an additional bond-bond interaction induced by the
continuous degrees of freedom;
specifically an expansion in the parameter $z\over T^{1\over 2}$
results in the expression
\(
f_{bb} = -{1\over 8T} z^2 \sigma_i \sigma_j \sigma_k \sigma_l
\langle (\theta_i - \theta_j)^2 (\theta_k - \theta_l)^2 \rangle_I
\label{fbb}
\)
where $(ij)$ and $(kl)$ are sites belonging to the same bonds and
$I$ refers to the irreducible correlator.  Conceptually, this
interaction (\ref{fbb}) is easier to understand as one between
{\em  ferromagnetic} bonds; if we define
\(
f_{ij} =   \left\{ \begin{array}{ll}
1 &\mbox{if $\sigma_i\sigma_j = 1$}\\
0 & \mbox{otherwise}
\end{array}
\right.
\label{fij}
\)
then we can write
\(
\sigma_i \sigma_j \sigma_k \sigma_l = 4f_{ij}f_{kl} - f_{ij} - f_{kl} - 1
\label{sig}
\)
which we insert in (\ref{fbb}).  Since the total number of FM bonds is
fixed in the ground-state manifold by the plaquette constraint, the terms
linear in $f_{ij}$ do not depend on the particular bond configuration
and thus can be neglected.  The bond-bond interaction can then be
written as
\(
f_{bb} = - {z^2 \over 32} \sum_{(ij)(kl)} {\cal A}_{(ij)(kl)} f_{(ij)}f_{(kl)}
\label{fbb2}
\)
where
\(
{\cal A}_{(ij)(kl)} = {1\over T} \langle (\theta_i - \theta_j)^2
(\theta_k - \theta_l)^2 \rangle_I
\label{Adef}
\)
can be expressed as a bilinear form
\(
{\cal A}_{(ij)(kl)} = 2g_{ik}^2 - 4g_{ik}g_{il} + ...
\label{bilinear}
\)
where
\(
g_{ik} \equiv {2\over T^{1\over2}} \langle (\theta_i - \theta_k)^2 \rangle
\label{gik}
\)
is {\em  independent} of temperature as $T \rightarrow 0$. Numerical evaluation
of the constant ${\cal A}_{(ij)(kl)}$ indicates that it is maximized by
FM bonds with a common vertex that form an angle of $2\pi\over 3$, and
decays rapidly with increasing bond-bond separation.  Therefore this
induced bond-bond interaction (\ref{fbb2}) favors the flat (111)
state which has the largest number of such bond configurations.
The free energy associated with the entropy of the fluctuating
surface $f_{Ising}\sim T$; since $f_{bb} \sim z^2$, at low
temperatures it enhances the height stiffness and results in the
sequence of transitions (paramagnet, liquid, magnet) discussed above
(Figure 6).

\section{The Canting Transition}

At low temperatures the spins described by the generalized model of the
previous section are energetically {\em  unstable} in any
collinear configuration, and the continuous variables $\theta_i$
acquire a small $(\sim z)$ non-zero average value at each site.
At $T=0$ a small-angle expansion of the energy
about the collinear
phase  results in the expression
\(
E({\theta}) = - {z\over 6} \sum_{ij} J_{ij} (\theta_{ij})^2
+ h\theta_i^2 + {\cal O}(\theta_{ij}^4)
\label{energy1}
\)
\[
J_{ij} = (1 + 3 \sigma_i \sigma_j)
\]
where $\theta_{ij} \equiv \theta_i - \theta_j$ and we have introduced
an additional nematic  field $h \cos 2\Theta$ in (\ref{ham})
to ``control'' the continuous degrees of freedom.  For the $100$
state each site has exactly two FM bonds, and the lowest eigenvalue
of the coupling matrix $J_{ij}$ is zero.  By contrast for the
$111$ phase, where each site has three FM bonds, the matrix
$J_{ij}$ has a minimum eigenvalue $\lambda_{min} = -9$.
The lowest eigenvalue of $J_{ij}$ associated with the typical
low-temperature state of the height variables is intermediate between
these two extreme cases and is always {\em  negative}.

This canting instability can be illustrated by considering the
angular deviation ($\theta_i$) of a spin at any site $i$ with
{\em  three} AFM bonds, where we maintain spins at all other sites
in their collinear positions.  From (\ref{energy1}) we find
that the Gaussian energy associated with such an
angular deviation is $E = -z(\theta_i)^2$; it has a
{\em  maximum} at $\theta_i =0$, and therefore the spin at site
$i$ is unstable to canting.  A higher-order expansion of the energy
(\ref{energy1}) about the collinear phase results in an effective
double-well potential associated with the continuous variables
$\theta_i$.  In the more general case where all the spins
are free to cant, the angular deviations on different sites are
correlated with a coupling determined by the matrix $J_{ij}$ in
(\ref{energy1}).  At intermediate temperatures where an effective
double-well potential develops, the spins are free to fluctuate
between its two minima. The ``freezing'' of each spin in one of these
potential wells will occur at a lower temperature, and the entropy
associated with this transition is $\ln 2$ per site.

The character of the transition is determined by the
coupling matrix $J_{ij}$ which is
the sum of two terms, one constant and the other determined by the
actual configuration of the surface; the latter can be viewed
as the interaction between the canting angles and the height degrees
of freedom.  The uniform contribution leads to an effective
$\phi^4$ theory which is in the Ising universality class.  This
fixed point behavior is stable against perturbative coupling
between the continuous and the discrete variables which,
by reasons of symmetry, must take the form
\(
F_{int} \propto (\nabla h)^2 \theta^2
\label{fint1}
\)
and is irrelevant at long wavelengths.  More specifically, an average of
(\ref{fint1})
over the height fluctuations shifts the transition temperature,
and a power-counting argument applied to the renormalization
of the $\phi^4$-interaction term
\(
\int \langle (\nabla h_0)^2 (\nabla h_r)^2 \rangle d^2 r  \sim {1\over L^2}
\label{fint2}
\)
indicates that it scales to zero at large distances.

In the small-angle expansion of the energy in (\ref{energy1}),
the interaction between the continuous and the height variables is significant,
and thus we cannot exclude the possibility of a first-order transition or
the presence of a new universality class.  However numerical studies of
this model, described below, indicate the stability of the Ising fixed
point.  As a result, we note that the formation of the canted state
has a feedback on the properties of the fluctuating surface.  Since the
coupling matrix $J_{ij}$ has its lowest minimum eigenvalue for the flat
$111$ phase, the angular variables have their lowest energy in this
configuration and thus modify the $T\rightarrow 0$ ground-state of the discrete
variables.  In the absence of an intervening first order transition,
we thus expect a continuous increase in the stiffness constant $\eta$
which preserves the sequence of three states (paramagnet, liquid, magnet)
of the Ising variables discussed above.  In Figure 7 we show a schematic
of the expected phase diagram; the relative positions of the canting
and the liquid phase boundaries
must be determined numerically.

\section{Numerical Diagnostics}

In order to check these ideas we have performed Monte Carlo simulations
on the model (\ref{int}) with $z=0.4$; this is the maximum value
for which no traces of $2\pi\over 3$ fluctuations were observed
in the low-temperature spin configurations ($z_c = 0.56$ for this model).
Triangular arrays of $6^2$, $12^2$, $24^2$,
$48^2$ and $96^2$ spins were sequentially cooled from a random
configuration at high temperatures with $2.5 \times 10^5$ spin flips/site
per temperature point, and thermal averages were taken over periods
of $2.5 \times 10^3$ updates, providing $10^2$ approximately
independent samples where the distribution of thermodynamic
variables was studied.
Some measurements (e.g. the surface stiffness coefficient) required longer runs
(total
number of flips per site $N_{tot} = 10^7$); in all cases the results were
checked for stability with respect to increased number of flips/site.
The
specific heat results are displayed in Figure 8.
At $T_{nem}=0.25$ there exists a nematic
transition into a phase with $<e^{2i\theta}>\neq 0$,
confirmed by the accompanying jump in the spin stiffness shown in
of Figure 9.
At $T_{cant}=0.03$
there is a sharp anomaly in the specific heat that increases with system size.
This data excludes
the possibility of an incipient
first-order transition at low temperatures and thus we expect the
sequence of transitions (paramagnet, liquid, magnet)
described schematically in Figures 6 and 7.

We associate the sharp structure in the specific heat data at low temperatures
(Figure 8)
with the canting instability.  A standard finite-size analysis
(see Figure 10) confirms that this transition is in the Ising universality
class.  Briefly for a finite system of size $L$ that is smaller
than the relevant correlation length $\xi$
the specific heat scales as
\(
c_v \sim L^{\alpha\over \nu} f(\epsilon L^{1\over \nu})
\label{fcv}
\)
where $\alpha$ and $\nu$ are the exponents associated with $c_v$ and $\xi$
in the limit $L \rightarrow \infty$, and $\epsilon \equiv \left( {T - T_c}\over
T_c \right)$
is the reduced temperature; here $f(x)$ is a universal function.
For the case of the $2D$ Ising ferromagnet $\alpha = 0$ and $\nu = 1$ so
that (\ref{fcv}) becomes
\(
c_v  =  A\log L + f(\epsilon L)
\label{nfcv}
\)
where
\(
\lim_{\epsilon L \rightarrow \infty} f(\epsilon L) = -A \log (\epsilon L)
\label{ffcv}
\)
ensuring that the $2D$ Ising results are recovered in the limit of infinite
system-size.
In Figure 10 we have displayed the
universal function $f(x)$ in (\ref{nfcv})
as a function of
$\log (\epsilon L)$ for five system sizes; the resulting
linear plot confirms
the Ising values of
$\alpha$ and $\nu$.
Furthermore we vary the coefficient $A$ in (\ref{nfcv})
to get the best fit to the numerical data;
its optimal value ($A = 0.50$) is equal to the slope
of the linear plot  in Figure 9,
consistent with (\ref{ffcv}). This coefficient $A$ is the amplitude associated
with the specific heat at the transition, and is consistent
with the known value ($A_{Ising}=0.53$) for the two-dimensional Ising
ferromagnetic model.\cite{huang}
A crude integration of the specific heat peak at $T_{cant}$
in Figure 8 leads to an estimated entropy $S \sim 0.85 \ln 2$;
both these  amplitude and the entropy results are consistent
with approximately one Ising degree of freedom/spin at
the transition.

In order to confirm that the $T_{cant}=0.03$ transition is due to the
presence of the continuous degrees of freedom, we apply
a nematic field
$h_{nem} = h \cos 2 \theta_i$
to suppress their thermal fluctuations.  The specific heat anomaly remains
sharp (Figure 11a) indicating that this field $h_{nem}$ is {\em  not} conjugate
to the order parameter; however it does affect the transition temperature
as shown in the resulting phase diagram in Figure 11b.
We can estimate the critical nematic field strength ($h_{crit}$) from the
quadratic form of  the energy;
it is the minimum value of $h$
which maintains the expression
(\ref{energy1})
positive definite.
For a typical state of Ising variables at $T=0.03$, we have
diagonalized the coupling matrix $J_{ij}$ and found that
$\lambda_{min} = -7.5$.  Thus the transition is completely
suppressed if $h_{nem} \ge 0.5$ in rough agreement with the numerical
results (Figure 11), where we were not able to follow the
phase line down to temperatures $T < 0.015$ due to problems of equilibration.

We can also estimate the canting transition temperatures using the quartic
terms in the
small-angle expansion of the energy (\ref{energy1}).  For this purpose we
consider again
the canting instability of a spin at site $i$ with three AFM bonds, where again
we maintain all its neighbors in their collinear positions.  The renormalized
energies
associated with AFM and FM bonds are
\(
\begin{array}{ll}
E^R_{AFM} &\approx \theta_{ij}^2 \left( {1\over 3} z + \langle\theta_{ij}^2
\rangle \right)\\
E^R_{FM}  &\approx \theta_{ij}^2 \left( -{2\over 3}z + \langle\theta_{ij}^2
\rangle \right)
\end{array}
\label{AFM-FM}
\)
so that the spin at site $i$ becomes unstable at
\(
{1\over 6} z \approx \langle \theta_{ij}^2 \rangle = {T\over \rho_s(T)}
\label{crit}
\)
where $\rho_s(T)$ is the spin stiffness; (\ref{crit}) is the defining equation
for the canting transition temperature.  Our measured quantities
\(
{T_c \over \rho_s(T_c)} \approx 0.06 \approx {z \over 6}
\label{us}
\)
are consistent with the expression (\ref{crit}).

In order to characterize the fluctuations of the Ising degrees of freedom,
we determine the height surface by measuring the relative height
fluctuations $\langle h_r ^2 \rangle \equiv \langle (h^2(r)\rangle -
\langle h(r)\rangle^2 $ between the middle and the edge of each array.  We
observe that for $T < T_{nem}$ the height fluctuations are described by the
$T=0$ fixed point of the pure Ising antiferromagnet; the logarithmic
behavior of the fluctuations, shown in Figure 12, is only robust
up to the average distance between height defects $(\xi \sim 3 \times 10^3$)
which exceeds all system sizes studied.  As expected, the
height stiffness $ \left( \langle h_r^2 \rangle =
{\ln L \over \eta \pi}\right)$
{\em  increases} in the approach to the canting transition; this behavior
is also displayed in the profile of the height fluctuations
as a function of decreasing temperature (Figure 13).
Unfortunately we cannot maintain the discrete degrees of freedom in equilibrium
to low enough temperatures to directly observe the presence of the
height liquid. However, since the specific heat data indicates no
first-order transition,  the surface stiffness must increase continuously;
furthermore we know, from our arguments about the canting instability,
that the surface will become spontaneously oriented with
$\eta > \eta_{c2}$
in the limit $T\rightarrow 0$. Our numerical results
confirm the expected enhancement of the surface stiffness coefficient
with decreasing temperature. We therefore infer
the presence of a
finite-temperature fluctuating surface with no defects;
its magnetic analogue is a spin liquid with power-law Ising
correlations.

Thus we observe  that in the vicinity of the canting transition, the surface
stiffness is enhanced.
We expect that for $T < T_{cant}$
$\eta$ will increase above its critical value ($\eta_{c1}$), resulting in
a finite-temperature height liquid on all length scales.
However on our length-scales of measurement even the Ising system
is a liquid, since at $T=0.03$ its correlation length
is several thousand lattice spacing, well beyond the largest system
sizes studied.  We have used the surface stiffness
to characterize the low-temperature behavior of the generalized model
and to distinguish it from its discrete counterpart.

The dynamics of the
resulting liquid, even at short length-scales,
is also very different from that of the pure Ising case
due to the interaction between its discrete and continuous
degrees of freedom.
In order to illustrate this feature,
we consider small changes to our fluctuating surfaces associated with
the addition/subtraction of an elementary block.  For the pure Ising case there
is
no energy cost associated with this process; however
it {\em  changes} the structure of the coupling matrix $J_{ij}$ in
(\ref{energy1}),
and results in a dramatic reconstruction of the continuous degrees of freedom.
As $T \rightarrow T_{cant}$ we expect that the canting
instability will appear first in the lowest eigenmodes
$J_{ij}$,
so that we can neglect all the others.
In Figure 14 we display a typical surface and its lowest eigenmode is
shown in Figure 15.  The subtraction of a cube not only rearranges
the energy level ordering but also modifies each state.  In order to
illustrate this property, we subtract a single block from the surface
displayed in Figure 14 at the location denoted by the arrow; we then
rediagonalize
the coupling matrix and display the eigenmode that has the
maximum overlap with that shown in Figure 15 .  The dramatic
reconstruction of the continuous degrees of freedom displayed
by comparison of Figures 15 and 16 requires some time;
in the transient
regime the energy of the angular degrees of freedom is high
implying the presence of macroscopic barriers associated
with this local process.

The preceeding discussion suggests a distribution of surface relaxation
times.\cite{pwa2,lm}  We thus measure the time-dependent overlap
\(
Q(t) = {1\over N} \sum_{i=1}^N (-1)^{h_i(t) - h_i(0)}
\label{overlap}
\)
between initial and final Ising states at a given temperature, probing
the system's exploration of phase space ; any feature in
$Q(t)$ indicates memory on the time-scale of those structure.\cite{overlaps}
Before
presenting our results, let us pause briefly to discuss
the possible scenarios.  In a high-temperature paramagnet (``gas'') the
exponential spatial and temporal spin correlations lead to an exponential
decay of the overlap.  By contrast, in a spin liquid the decay of $Q(t)$
is power-law; this is the situation for the $T \rightarrow 0$ Ising
triangular antiferromagnet, and is maintained for this discrete case
at any finite temperature up to the length-scale defined by the defect-defect
separation.  In a glass the functional form of the overlap decay is
yet slower than power-law due to the broad distribution of relaxation times.
A slow decay
of the overlap on a time-scale $\tau$ indicates the presence of barriers
of height $\sim T \ln \tau$, and the magnitude of $Q(\tau)$ measures the number
of
metastable states separated by these barriers.  For example, the
Sherrington-Kirkpatrick (SK) spin glass, where the interaction
is infinite-range, has a minimum correlation volume that scales with
the system-size ($L$);
the associated overlap decay indicates that both the barrier number
and their maximum height scale as powers of $L$.
By contrast, the Edwards-Anderson spin glass with its short-ranged
interaction, displays a broad distribution of length-scales;
the associated barrier distribution is continuous and only limited
by $L$, where the number of barriers with maximum height
scales as $L^\alpha$ where $\alpha > 0$.

In Figures 17 and 18 we show our results for the time-dependent overlap of the
generalized
model above and below the canting transition; the measurements are averaged
over different time-scales.  In Figure 17 we observe that the
short-time behavior of $[Q(t)]^2$ is qualitatively similar for
$T > T_{cant}$ and $T < T_{cant}$; both display similar features
at these time-scales but are different from the pure Ising case where
this $[Q(t)]^2$ is completely flat (dotted line in Figure 17). At longer
times (Figure 18) only the overlap measured at $T < T_{cant}$ has discernible
structure; $[Q(t)]^2$ for $T > T_{cant}$ on these time-scales is featureless.
Figure 18a indicates that for $T < T_{cant}$ the system is making transitions
between a few metastable states separated by large barriers of similar
height; their size is proportional to the logarithm of the
pseudo-period of the observed undulations.
This data also suggests that there is a minimum correlation volume associated
with this generalized model; presumbably it is related to the number of sites
involved in the reconstruction of the $\theta_i$ variables following
an elementary process.
A study of the identical
overlap for a larger system-size is necessary to determine the finite-size
scaling of the maximum barrier height and number; in particular
we would like to investigate whether the system becomes ``localized''
in phase space thereby breaking ergodicity.  The results displayed in
Figure 18b indicate that the barrier heights do increase for larger $L$;
however
\(
\lim_{t \rightarrow \infty} [Q(t)]^2 = \lim_{t\rightarrow \infty}
[Q_{Ising}(t)]^2
\label{isingov}
\)
so that the number of metastable states does {\em  not} increase with
system-size and the minimum correlation volume remains finite.  Thus, though
the system retains some long-time memory
of its initial Ising state, it is a ``slow'' viscous liquid and
{\em  not} a glass; this is because in the thermodynamic limit
$(L \rightarrow \infty)$ it can explore almost all of phase space and
thus is effectively ergodic.

\section{Discussion}

In summary we have constructed a frustrated spin model whose discrete degrees
of freedom acquire power-law correlations at finite temperatures.  There is a
true thermodynamic transition into this ``spin liquid'', which exists as
an
intermediate phase separating the paramagnetic (``gas'') and the magnetic
(``solid'') states.
This spin liquid has an extensive zero-point entropy and is
the finite-temperature analogue of the $T\rightarrow 0$ phase of
the Ising antiferromagnet on the triangular lattice.  This highly
degenerate ground-state manifold is conveniently parametrized by an
undulating surface, whose fluctuations on large length-scales are characterized
by a surface stiffness constant.  For the purely discrete model, the mapping
between the spin and the membrane problem is no longer uniquely defined at
any finite temperature due to the presence of singularities in the surface.

In this paper we have constructed an $xy$ spin model on the triangular
lattice that has both continuous and Ising low-energy excitations; the
ground-state configurations of these discrete variables are isomorphic to those
of the pure Ising antiferromagnet above.  In the generalized model, the
interaction between the $xy$ and the Ising degrees of freedom results
in the continuous enhancement of the surface stiffness with decreasing
temperature.  As a result, the defects are bound at finite temperatures
and the surface mapping becomes well-defined. For very low temperatures
the stiffness coefficient becomes sufficiently large to spontaneously
orient the surface,
and the ground-state is unique.

We thus expect a sequence of three phases (paramagnet, liquid, magnet)
associated with the discrete variables as a function of decreasing
temperature.  There is also a canting instability where the
angular deviations $\theta_i$ acquire a non-zero average at each site;
the locations of these canting and liquid phase boundaries must
be determined numerically. Similarly the possibility of an intervening
first-order transition cannot  be excluded analytically.  Therefore
we have performed Monte Carlo simulations on the generalized $xy$ model
described here.  At low temperatures ($T_{cant}=0.03$ for $z=0.4$)
we observe a sharp anomaly in the specific heat which we attribute
to the canting transition; a finite-size scaling analysis of these results
indicates that it is in the Ising universality class.  Application of
a nematic field suppresses $T_{cant}$, confirming that it is due to the
continuous degrees of freedom.  The resulting phase line and the measured
values
of $T_{cant}$ and $\rho_s(T = T_{cant}$) are all consistent with that
expected from a small-angle expansion of the energy.  Measurement of the
surface
stiffness coefficient confirms its enhancement with decreasing temperature,
though unfortunately equilibration problems prevent the direct observation
of the height liquid when $\eta > \eta_{c1}$.
The interaction between the discrete and the continuous variables
also leads to the presence of macroscopic barriers associated with
local processes.  In order to probe the systems exploration of
phase space, we measure the overlap between the final and initial Ising
states averaged over different time-scales; we find that
though the system has some long-term memory it remains effectively
ergodic in the thermodynamic limit.

More generally, we would like to emphasize that the classical liquid
discussed here is {\em  not} an anomalous result restricted to the specific
Hamiltonian constructed above.  In the broadest terms, the stability of such
a finite-temperature phase with power-law discrete spin correlations
could emerge from any model with the following crucial features:
\begin{quote}
{\sl (i)}  At low temperatures it should have {\em  both} discrete and
continuous spin excitations \\
{\sl (ii)} Its discrete degrees of freedom retain an extensive
entropy\cite{pb} at $T \rightarrow 0$ and, in all known examples
in finite-dimensions, the associated
ground-state manifold is conveniently described by a freely fluctuating
surface \\
{\sl (iii)} The interaction between the discrete and the continuous degrees of
freedom should enhance the  stiffness constant of the associated surface
\end{quote}
We note, for example, that an $xy$ model on the kagom\'e lattice
satisfies all three of these conditions; furthermore since
$\eta_{Potts} = \eta_{c1}$  we expect that a weak interaction between
the continuous and discrete degrees of freedom should be sufficient to
drive the system into a liquid phase.

The enhanced stiffness constant associated with the model specified by
criteria $(i)$, $(ii)$ and $(iii)$ not only affects the equilibrium
properties of the underlying discrete degrees of freedom, but modifies
their dynamics as well. In particular a local process, which in the purely
discrete model has a negligible energy cost, can now involve significant
rearrangement of the continuous degrees of freedom; this surface
reconstruction is not instantaneous and thus there will be a
transient regime where
the surface energy is large.  The distribution of the associated
barriers can be
probed by a study of the overlap decay between final and initial discrete
states. For the specific model studied in this paper, we find that
there is a large minimum correlation volume that remains finite in
the thermodynamic limit resulting in a viscous spin liquid
that effectively retains its ergodicity.
Though the initial Hamiltonian
has purely local couplings,  the interplay between
the Ising and the $xy$ degrees of freedom  generates
long-range interactions between discrete spins on a
length-scale that is large but finite in the limit $L \rightarrow \infty$.
The development of a glass requires the generation of an
infinite-range correlation; this may be difficult to achieve in two
dimensions due to the large phase space available to long-wavelengh
spin
fluctuations.
It is difficult to construct an analogous model in higher dimensions
since the tendency towards magnetic ordering is larger than in $d=2$,
and thus it will be difficult to find a discrete system that retains
its extensive entropy in the limit $T \rightarrow 0$ and thus to
satisfy $(ii)$.
However, if found, we expect it to have
still slower dynamics and even possible glassiness in the absence of
disorder.

We are very grateful to B. Doucot and L. Levitov for sharing their
ideas and insights with us as this work progressed, and to
M.V. Feigelman for a critical reading of the manuscript.
The numerical work was performed on an NEC SX-3 at the HNSX Supercomputer
Center (Woodlands, Texas), and we thank N. Trouillier for
assistance with the code. P. Coleman
acknowledges financial support from NSF Grant No. DMR-89-13692.

\endreferences

\begin{figure}
\caption{A section of the $(111)$ ground-state configuration
of the Ising triangular
model a) before and b) after the
$\pi$-flip of a spin (bold) with exactly three AFM bonds (solid
lines); all other circled spins can be similarly
reversed at zero energy cost resulting in
a lower bound for
the $T \rightarrow 0$ entropy of  $S \ge {N\over 3} \ln 2$.}
\end{figure}

\begin{figure}
\caption{A section of the $(100)$ ground-state configuration of
the Ising triangular model a) before and b) after the
$\pi$-flip of a spin (bold) with four AFM bonds (solid lines);
this process creates two ``bad'' triangles (shaded) whose subsequent separation
by an additional spin-flip, illustrated in c), does not further
violate the plaquette constraint.}
\end{figure}

\begin{figure}
\caption{A schematic illustrating that the energy cost associated
with two defects in the Ising triangular model
is independent of their relative position; here only the AFM bonds are
displayed,
and the dotted lines refer to the bonds ``flipped''
in moving a defect from its location in a) to that in b).}
\end{figure}

\begin{figure}
\caption{A section of a typical ground-state Ising configuration (a) with its
corresponding rhombi tiling (b) and its associated surface
representation (c) where the heights ($h_i$) are indicated.}
\end{figure}

\begin{figure}
\caption{The height differences around a closed loop (bold line with arrows)
in the
a) the absence
and
b) the presence
of a height defect (dotted lines indicate the bonds ``flipped''
to create the defect pair); note that this loop sum is zero if the number of
defects
enclosed is even but is a multiple of six if it is odd.}
\end{figure}

\begin{figure}
\caption{Schematic of the sequence of paramagnet-liquid-magnet transitions
expected as a function of increasing height stiffness constant.}
\end{figure}

\begin{figure}
\caption{Proposed phase diagram for the generalized spin model described in
the text where C, L and S refer to the canted, liquid and solid phases
repectively and the dotted line indicates a first-order transition     .}
\end{figure}

\begin{figure}
\caption{Specific heat results for the $z=0.4$ model described in the
text for arrays of $6^2$,$12^2$,
$24^2$, $48^2$ and $96^2$ spins.}
\end{figure}

\begin{figure}
\caption{Spin Stiffness Results for the $z=0.4$ model described in the
text for arrays of $6^2$,$12^2$,
$24^2$, $48^2$ and $96^2$ spins.}
\end{figure}

\begin{figure}
\caption{Finite-size Scaling Analysis for the Specific Heat Data
$T<T_c$ where $T_c = 0.03$;
the linear behavior of this plot confirms that the associated canting
transition is in the Ising universality class (similar results
are obtained for the $T>T_c$ case).}
\end{figure}

\begin{figure}
\caption{a) Specific heat data for $h_{nem} = 0.0,0.3$ indicating that
the anomaly remains sharp despite the suppression of $T_c$
and b) Phase diagram of $h_{nem}$ vs $T_{Ising}$ where $h_{nem}$
is an applied nematic field (see text).}
\end{figure}

\begin{figure}
\caption{$<h_r^2>$ vs. $\ln L$ where the
height fluctuations $\langle h_r^2 \rangle \equiv
\langle h^2(r)\rangle - \langle h(r)\rangle^2$ are related to
the height stiffness ($\eta$) and the array length ($L$)
by the expression $<h_r^2> = {\ln L \over \eta\pi}$.}
\end{figure}

\begin{figure}
\caption{Profile of height fluctuations across the diagonal
of a $96 \times 96$ array as a function of decreasing temperature.}
\end{figure}

\begin{figure}
\caption{A snapshot of a ``typical'' surface in the vicinity of
the canting transition; the arrow points to one location where the
addition/subtraction of an elementary cube can occur.}
\end{figure}

\begin{figure}
\caption{The lowest eigenmode ($|\Psi|$) of the  ``typical'' state in
Fig. 14 in the vicinity of the canting transition}
\end{figure}

\begin{figure}
\caption{
An eigenmode
corresponding to a state after the subtraction of
an elementary building block in Fig. 14 at the location
denoted by the arrow; the eigenmode displayed here has a maximal
overlap with the original one shown in Fig. 15.}
\end{figure}

\begin{figure}
\caption{The time-dependent height overlap
$[Q(t)]^2 =
\left({1\over N_i} \sum_i (-1)^{\{h_i(t) - h_i(0)\}}\right)^2$
averaged over a) 10 and b) 100 Monte Carlo steps for a $24^2$ array; its
behavior for $T>T_c$
and $T<T_c$ is qualitatively similar.}
\end{figure}

\begin{figure}
\caption{The time-dependent height overlap
$[Q(t)]^2 =
\left({1\over N_i} \sum_i (-1)^{\{h_i(t) - h_i(0)\}}\right)^2$
averaged over 1000 Monte Carlo steps for a a) $24^2$
and a b) $48^2$ array;
the pseudo-periodic structure implies the presence of large
barriers whose height but {\sl not} number increase with system
size and thus the system is effectively ergodic in the thermodynamic
limit.}
\end{figure}


\begin{references}
\bibitem{mc}We determined the magnitude of this prefactor using a Monte
Carlo simulation.
\end{references}
\end{document}